\documentclass[%
 reprint,
 amsmath,amssymb,
 aps,
]{revtex4-2}

\usepackage{graphicx}
\usepackage{dcolumn}
\usepackage{caption}
\usepackage{subcaption}
\usepackage{amsmath}
\usepackage{bm}
\usepackage{gensymb}

\begin{document}

\preprint{APS/123-QED}

\title{Revisiting the Longitudinal Development of Electromagnetic Air Showers: Analytical Improvements to the Greisen Formalism with Zenith Angle Dependence}
\author{Sebastián Mendizabal$^{a}$}
\author{Nicolás Viaux M.$^{a,d}$}
\author{Sebastián Tapia$^{a,c}$}
\author{Raquel Pezoa R.$^{b,c}$}
\author{Barbara Gutiérrez$^{a}$}
\author{Constanza Valdivieso$^{a}$}

\affiliation{$^{a}$Departamento de Física, Universidad Técnica Federico Santa María, Valparaíso, Chile}
\affiliation{$^{b}$Departamento de Informática, Universidad Técnica Federico Santa María, Valparaíso; Chile}
\affiliation{$^{c}$Centro Científico Tecnológico de Valparaíso (CCTVal), Universidad Técnica Federico Santa María, Valparaíso, Chile}
\affiliation{$^{d}$Millennium Institute for Subatomic Physics at the High Energy Frontier (SAPHIR), Santiago, Chile}

\date{\today}

\begin{abstract}

We present a new analytical approach to the longitudinal development of electromagnetic air showers, offering improvements to the classical Greisen formalism. We introduce a novel profile for the slope function $\lambda_1(s)$ that achieves an agreement less than $0.75\%$ with the original $\lambda_1$ for shower age parameter $s$ between $0.3 < s < 1.4$, where $s$ represents the stage of shower development. Our new formalism provides an improved representation of shower evolution, particularly near and beyond the shower maximum. In addition, we derive a complete expression for the number of particles $N(t)$. Our implementation includes the zenithal angle dependence on the number of particles at the detector level at high altitudes, making it particularly useful for high-altitude observatories. This expression is suitable for implementing air shower simulation tool fitting procedures over a wide range of energies and geometries. Our analysis suggests that the proposed new formalism may provide better agreement with the expected evolution of particle numbers compared to the traditional Greisen formulation.
\end{abstract}

\maketitle


\section{Introduction}

The study of cosmic ray air showers is fundamental to our understanding of high-energy astrophysical phenomena. When a high-energy cosmic ray particle interacts with the Earth's atmosphere, it initiates a cascade of secondary particles — an extensive air shower (EAS) — whose development is shaped by the energy and type of the primary particle, as well as the atmospheric density profile \cite{gaisser2003cosmic}. A central challenge in astroparticle physics is to model the number and type of particles that reach a given observational altitude, especially for ground- or balloon-based detectors. This modeling is crucial for interpreting data from cosmic ray observatories and understanding the properties of the primary particles.

A key milestone in the theoretical description of electromagnetic showers is the Greisen profile, derived from solutions to the cascade equations in the Rossi-Greisen approximation \cite{greisen1956cosmic}. This formulation led to the expression for the number of electrons as a function of depth \( t \), which is defined as:
\begin{equation}
t = \frac{X}{X_0},
\end{equation}
where $X$ is the atmospheric depth and $X_0$ is the radiation length. This formulation led to the expression for the number of electrons that arrives to the ground at certain altitude produced in the  cosmic ray shower:
\begin{equation}
N(t) \sim \frac{\epsilon}{\sqrt{\beta_0}} \, \exp\left[t\left(1 - \frac{3}{2} \ln s\right)\right],
\end{equation}
where \( \beta_0 = \ln(E_0 / \epsilon_c) \) is the logarithm of the ratio between the primary energy \( E_0 \) and the critical energy \( \epsilon_c \), which is the energy at which ionization losses equal radiative losses in the atmosphere. For electrons in air, its value is approximately \( \epsilon_c \approx 81 \, \mathrm{MeV} \) \cite{gaisser2003cosmic}. The parameter \( s \) is known as the shower age (see for example \cite{Lipari:2008td,Giller:2004cf,Nerling:2005fj,Gora:2005sq} for some studies on the concept of shower age), and it comes from the relation with the parameter $t$ (shown later), representing the stage of cascade development. The co efficient \( \epsilon \) arises from a saddle-point approximation in Mellin space and is evaluated at the shower maximum. Its precise value depends on the second derivative of the profile \cite{greisen1956cosmic}.

In this paper, we present a new analytical approach to improve the Greisen formalism by developing a novel profile for the slope functions $\lambda_1(s)$ and $\lambda_2(s)$. These functions, originally introduced by Rossi and Greisen \cite{Rossi1941}, play a fundamental role in the mathematical description of cosmic ray shower development.

The behavior of these lambda functions varies with the shower age parameter $s$, which characterizes the developmental stage of the cascade: for $s < 1$, the shower is in its development phase; at $s = 1$, it reaches maximum development with the highest number of particles; and for $s > 1$, it enters the absorption phase. These functions appear in key equations describing differential and integral spectra of electrons and photons, and form the mathematical foundation for calculating observable properties such as track length, center of gravity, and longitudinal spread of particles in the cascade.

Our new formulation achieves an agreement of less than $0.75\%$ with the original $\lambda_1$ for shower age parameter $s$ between $0.3 < s < 1.4$, providing significantly improved accuracy in this critical range for experimental applications. This enhancement is particularly valuable for modern cascade modeling, as these functions remain central to predicting particle numbers at different atmospheric depths.

Furthermore, we extend our analysis to include the zenith angle dependence on the number of particles at ground level at high altitudes, making our approach particularly valuable for high-altitude observatories. This comprehensive treatment allows for more accurate predictions across a wider range of observation conditions than previously available analytical models, significantly expanding the scope of application of the classical formalism.

\section{Theoretical Framework}

\subsection{The Greisen Formalism}

A key analytical tool in the modeling of the longitudinal development of air showers is the slope function $\lambda_1(s)$, introduced in the framework of cascade theory by Greisen \cite{greisen1956cosmic}. This function is defined as the logarithmic derivative of the number of particles with respect to atmospheric depth:
\begin{equation}
\lambda_1(s) = \frac{1}{N(t)} \frac{dN(t)}{dt},
\end{equation}
where $t$ is the atmospheric depth in units of radiation length, and $s$ is the shower age parameter representing the stage of cascade development. This differential relation leads to a general solution for the particle number:
\begin{equation}
N(t) = N_0 \exp\left( \int \lambda_1(s) \, dt \right),
\end{equation}
where $N_0$ is an overall normalization constant.

In the specific case of the Greisen profile, the slope function is given by (see \cite{Schiel:2006vf} for a full derivation):
\begin{equation}
\lambda^{1}_{\text{Greisen}}(s) = \frac{1}{2} (s - 1 - 3 \ln s),
\end{equation}
and the relation between shower age $s$ and atmospheric depth $t$ is:
\begin{equation}
s = \frac{3t}{t + \ln(E_0 / \epsilon_c)},
\end{equation}
which leads to the well-known expression for the number of electrons:
\begin{equation}
N(t) = \frac{\epsilon}{\sqrt{\ln(E_0/\epsilon_c)}} \exp\left[t\left(1 - \frac{3}{2} \ln s\right)\right].
\end{equation}
The numerical value of \(\epsilon\) will be discussed later. This formulation has been widely used in air shower modeling due to its analytical simplicity. However, it has limitations in accurately representing shower development across the full range of the shower age parameter, particularly for values of $s < 1$.

\subsection{Proposed Modified Slope Function}

Different profiles have been used to describe the evolution of the shower, which can be referred as intermediate models, as shown in \cite{Montanus:2011hg,Matthews:2009zg}, these models are very similar to each other but not as realistic as the Greisen one. We propose a modified slope function, designed to match more closely the numerical behavior of the shower growth rate across the shower development range $0.3 < s < 1.4$. The new profile $\lambda_{MV}^{1}(s)$ (MV for Mendizabal-Viaux) satisfies the essential boundary conditions $\lambda_{MV}^{1}(1)=0$ and $\lambda_{MV}^{1'}(1)=-1$, which correspond to the shower maximum ($s=1$).

While the Greisen profile describes $\lambda_1(s)$ quite well for values of $s>1$, it requires adjustment for lower values of $s$. Our approach introduces an exponential term that specifically improves the profile in this critical range:
\begin{multline}
\lambda^{1}_{MV}(s) = \frac{1}{2A}\big(1.215s - 1.215 - 3.2\ln(s) \\
+ e^{1 - s/0.1} - e^{1 - 1/0.1}\big).
\end{multline}

The derivative of this function is given by:
\begin{equation}
\lambda^{1'}_{MV}(s) =\frac{1}{A}\left(\frac{1.215}{2} - \frac{3.2}{2s} - \frac{1}{0.2} e^{1 - s/0.1}\right).
\end{equation}

The coefficient $A$ will guarantee that the derivative of the profile evaluated at $s=1$ is $\lambda^{1'}_{MV}(1)=-1$, for our case $A\approx0.99312$. The parameters in our modified function have been carefully adjusted to optimize agreement with numerical calculations. The coefficients 1.215 and 3.2 represent refined values of the original Greisen parameters (1 and 3), while the exponential term provides the necessary correction for early shower development stages.

\begin{figure}[htbp]
\centering
\includegraphics[width=\columnwidth]{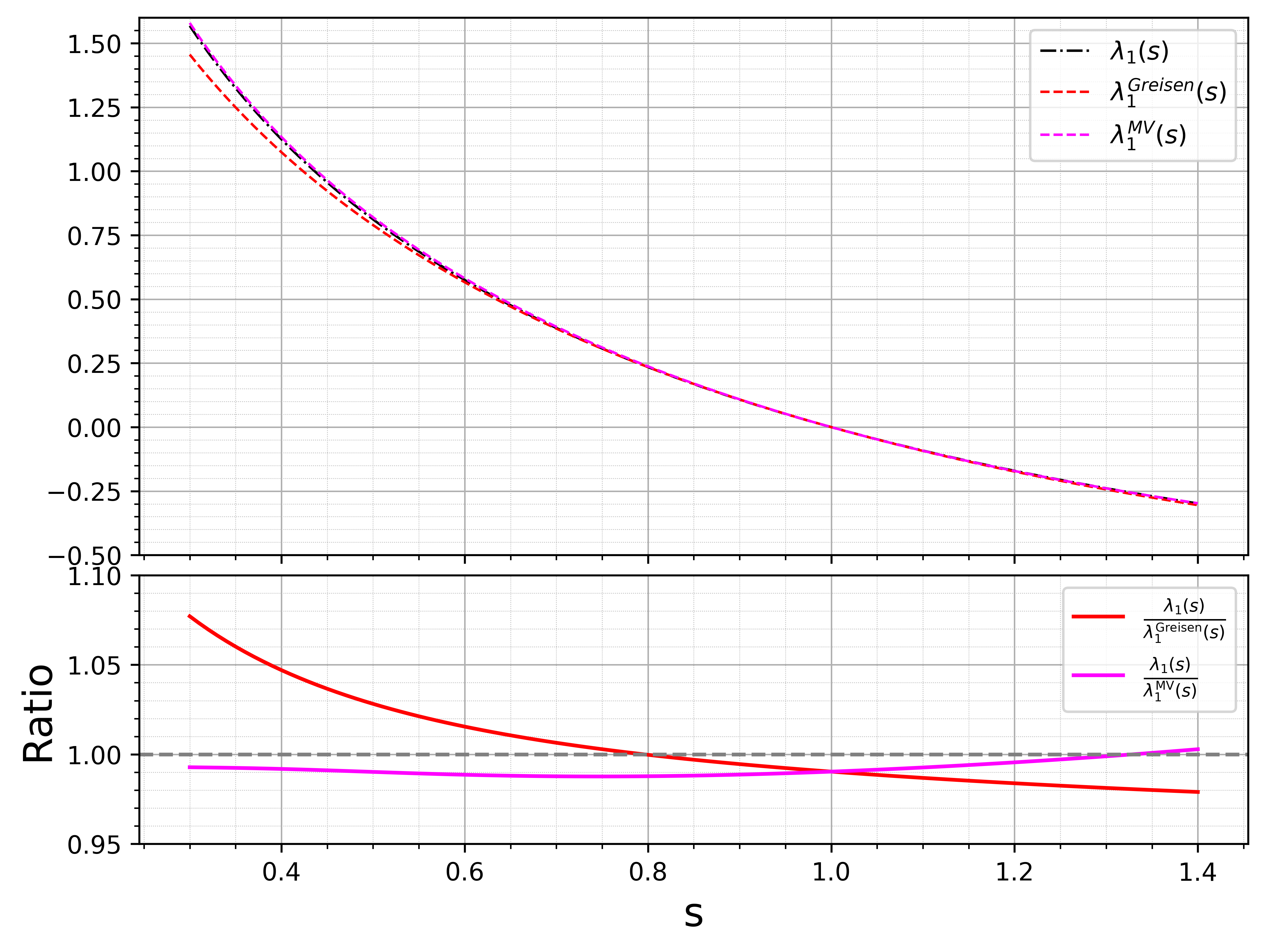}
\caption{Comparison of $\lambda_1$ (direct calculation, eq. 5.34 of \cite{gaisser2003cosmic}), $\lambda_1$ approximation of Greisen, and our new model $\lambda^1_{MV}$. The graph demonstrates that our model and the direct calculation across the shower age agrees in the range $0.3 < s < 1.4$ is $<0.75\%$.}
\label{fig:lambda1}
\end{figure}

As shown in Figure \ref{fig:lambda1}, our modified function achieves excellent agreement with the direct calculation of $\lambda_1$ across the critical shower age range, representing a significant improvement over the original Greisen approximation.

\subsection{Relationship Between Shower Age and Atmospheric Depth}

From our modified slope function, we can derive the relationship between shower age $s$ and atmospheric depth $t$ through the following approximation:
\begin{equation}\label{tdefinition}
t = -\frac{1}{\lambda^{1'}_{MV}(s)}\left(\beta_0 - \frac{1}{s}\right) \approx -\frac{\beta_0}{\lambda^{1'}_{MV}(s)},
\end{equation}
where $\beta_0 = \ln(E_0/\epsilon_c)\gg1/s$. With our new profile, this relationship becomes:
\begin{equation}
    t=\frac{2A\beta_0}{1.215-\frac{3.2}{s}-\frac{\exp(1-s/0.1)}{0.1}}.
\end{equation}

This formulation allows us to accurately map between shower age and atmospheric depth, which is essential for practical applications in air shower modeling.

\subsection{Derivation of the Particle Number Expression}

Using the relationship between $s$ and $t$, we can change variables in the integral for the particle number:
\begin{eqnarray}
N(t) &=&N_0\exp\left( \int \lambda^1_{MV}(s) \, dt \right),\nonumber \\ 
&=&N_0 \exp\left( \beta_0 \int \frac{\lambda^1_{MV}(s)\lambda ^{1''}_{MV}(s)}{(\lambda ^{1'}_{MV}(s))^2} \, ds \right).\label{ntfullapprox}
\end{eqnarray}

By integrating by parts and using the relation of $s$ and $t$ given in Equation (\ref{tdefinition}), we obtain:
\begin{equation}\label{lambdarelation}
    \int \lambda^1_{MV}(s) \, dt=\lambda^1_{MV}(s)t+s\beta_0,
\end{equation}

The normalization constant $N_0$ is evaluated at shower maximum, where $s=1$ which implies $t_{max}=\beta_0=\ln({E_0/\epsilon_c})$. This maximum is normally calculated at the zenith angle $0\degree$. For a different angle, this result changes by a factor $t\rightarrow t/C$, where for a flat earth approximation (under 60°) we get $C=\cos\theta$. The maximum at an angle different from the zenith can be written as $t_{max}=\beta_0\cos\theta$, we will discuss the angular dependence later in this work.

The approximate solution, as shown in \cite{greisen1956cosmic}, of the differential equation for the number of particles at different $t$ with initial energy $E_0$ is 
\begin{eqnarray}\label{fullsolution}
    N(E_0,t)&=&\frac{1}{\sqrt{2\pi}}\bigg[\bigg(\frac{E_0}{\epsilon_c}\bigg)^s\frac{ K_1(s,-s)}{s}\\
    &&\hspace{1cm}\times\frac{G_{\gamma\rightarrow e}}{\sqrt{\lambda_{MV}^{1''}(s)t+1/s^2}}e^{\lambda_{MV}^{1}(s)t}\bigg]\ ,\nonumber\\
   &\approx& \frac{1}{\sqrt{2\pi}}\bigg[\bigg(\frac{E_0}{\epsilon_c}\bigg)^s\frac{ K_1(s,-s)}{s}\frac{G_{\gamma\rightarrow e}}{\sqrt{\lambda_{MV}^{1''}(s)t}}e^{\lambda_{MV}^{1}(s)t}\bigg].\nonumber
\end{eqnarray}

Equation (\ref{fullsolution}) represents an approximation of the full solution, where an additional term involving $\exp[\lambda_{MV}^2(s)t]$ may be included. However, since $\lambda_{MV}^2(s)$ is always negative and $|\lambda_{MV}^2(s)|>|\lambda_{MV}^1(s)|$ this term becomes negligible for large $t$. At shower maximum, this approximation simplifies to:
\begin{equation}
N_{\text{max}} = \frac{1}{\sqrt{2\pi}} \left( \frac{E_0}{\epsilon_c} \right) K_1(1, -1) \frac{G_{\gamma \rightarrow e}}{\sqrt{\lambda_{MV}^{1''}(1) \beta_0}},
\end{equation}
where we have used the fact that $\beta_0 \gg s$. Comparing the last equation with (\ref{ntfullapprox}), leads to the normalization:
\begin{equation}
N_0 = N_{\text{max}} \left(\frac{\epsilon_c}{E_0}\right).
\end{equation}
Therefore, the final expression for the longitudinal profile becomes:
\begin{equation}\label{approxsolution}
N(E_0,t) =\frac{0.313}{\sqrt{\beta_0}}\exp\left( \lambda^1_{MV}(s)t+s\beta_0 \right),
\end{equation}
where the prefactor 0.313 results from evaluating the full normalization expression:
\begin{equation}\label{prefactorepsilon}
\frac{1}{\sqrt{2\pi}}\frac{K_1(1,-1)G_{\gamma\rightarrow e}}{\sqrt{\lambda_{MV}^{1''}(1)}} \approx 0.313,
\end{equation}
where we have used $K_1(1,-1)= 2.3082$ and $G_{\gamma\rightarrow e}(s=1)=G_{e\rightarrow e}(s=1)=0.4332$ \cite{Rossi1941,Lipari:2008td}. We note that by considering the second derivative of the Greisen profile in equation  (\ref{prefactorepsilon}) we obtain $\epsilon=0.326$ that differs from the widely used $\epsilon=0.31$, shown in \cite{gaisser2003cosmic}.  

Notice that by comparing the full solution in Equation (\ref{fullsolution}) with Equation (\ref{approxsolution}) and using the relation:
\begin{equation}
\left(\frac{E_0}{\epsilon_c}\right)^{s}=e^{s\beta_0},
\end{equation}
we arrive at the same solution as given by Equation (\ref{lambdarelation}).

\subsection{Derivation of $\lambda_2(s)$}

To derive the second root of the power-law solution to the cascade equations,  $\lambda_2(s)$, from our improved $\lambda^1_{MV}(s)$ function, we consider the general expressions for $\lambda_{1,2}$  in terms of the auxiliary functions, following  \cite{gaisser2003cosmic}
\begin{eqnarray}\label{lambassoluciones}
    \lambda_{1,2}=&&-\frac{1}{2}(A(s)+\sigma_0)\nonumber\\&&\pm\frac{1}{2}\sqrt{(A(s)-\sigma_0)^2+4B(s)C(s)},
\end{eqnarray}
where $\sigma_0$ comes from the total probability of pair production and is given by 
\begin{equation}
    \sigma_0=\frac{7}{9}-\frac{b}{3},
\end{equation}
and the parameter $b=(18\ln[183/Z^{1/3}])^{-1}=0.0122$, with $Z=7.4$ for air \cite{agostinelli2003geant4}. The auxiliary functions $B(s)$ and $C(s)$ can be obtained numerically, as they do not exhibit the divergence seen in the case of $A(s)$ \cite{gaisser2003cosmic}. We can rearrange the solutions to obtain a semi-analytical expression for $\lambda^2_{MV}(s)$ in terms of $\lambda^1_{MV}(s)$ by writing:
\begin{equation}
    A(s)=\frac{B(s)C(s)-\lambda_{MV}^1(s)(\lambda_{MV}^1(s)+\sigma_0)}{\lambda_{MV}^1(s)}.
\end{equation}
This allows us to compare our new expression for $\lambda^2_{MV}(s)$ in equation (\ref{lambassoluciones}) with the use of $\lambda^1_{MV}(s)$, providing an additional validation of our approach.

\begin{figure}[htbp]
\centering
\includegraphics[width=\columnwidth]{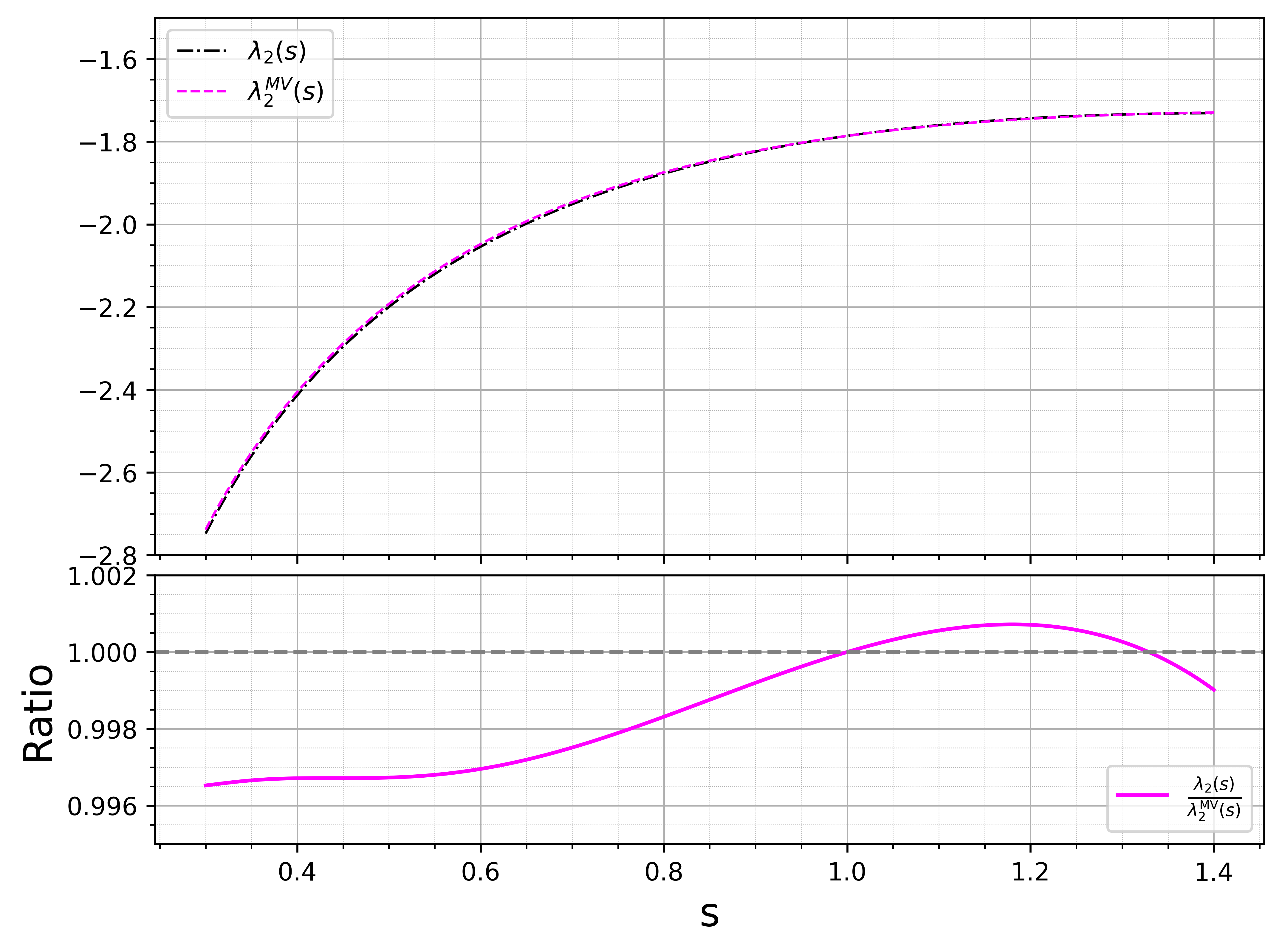}
\caption{Comparison between $\lambda_2$ (direct calculation, eq. 5.35 of \cite{gaisser2003cosmic}) and $\lambda_{MV}^{2}$ that was calculated from our improved $\lambda_{MV}^{1}$. The close agreement validates the consistency of our approach.}
\label{fig:lambda2}
\end{figure}

Figure \ref{fig:lambda2} shows the comparison between the directly calculated $\lambda_2$ function and the $\lambda^2_{MV}$ derived from our improved $\lambda^1_{MV}$ model. The excellent agreement further validates our approach and demonstrates the internal consistency of our formalism.

\subsection{Angular Dependence}

For cosmic rays arriving at zenith angles different from $0\degree$, we need to generalize our formalism to account for the increased atmospheric path length. This can be accomplished by considering how the atmospheric depth $t$ changes with the zenith angle.

For inclined showers, the atmospheric depth traversed increases by a factor of approximately $1/\cos\theta$ for zenith angles below $60\degree$ (assuming a flat Earth approximation). This affects the shower development and the position of the shower maximum, which becomes:
\begin{equation}
t_{max}=\beta_0\cos\theta.
\end{equation}
The number of particles at a given atmospheric depth for an inclined shower can be calculated by applying the transformation:
\begin{equation}
t \rightarrow \frac{t}{\cos\theta}
\end{equation}
to our expression for $N(E_0,t)$. This yields:
\begin{equation}
N(E_0,t,\theta) = \frac{0.313}{\sqrt{\beta_0\cos\theta}}\exp\left(\lambda^1_{MV}(s)\frac{t}{\cos\theta}+s\beta_0\right),
\end{equation}
where $s$ is now related to $t$ through:
\begin{equation}
t = -\frac{\beta_0\cos\theta}{\lambda^{1'}_{MV}(s)},
\end{equation}


This angular dependence is significant, where the atmospheric overburden is smaller and the shower development stage observed can vary significantly with zenith angle. Our improved $\lambda_1^{MV}(s)$ function maintains its accuracy across different zenith angles, making it valuable for modeling inclined showers at high-altitude observatories.

\begin{figure}[htbp]
\centering
\includegraphics[width=\columnwidth]{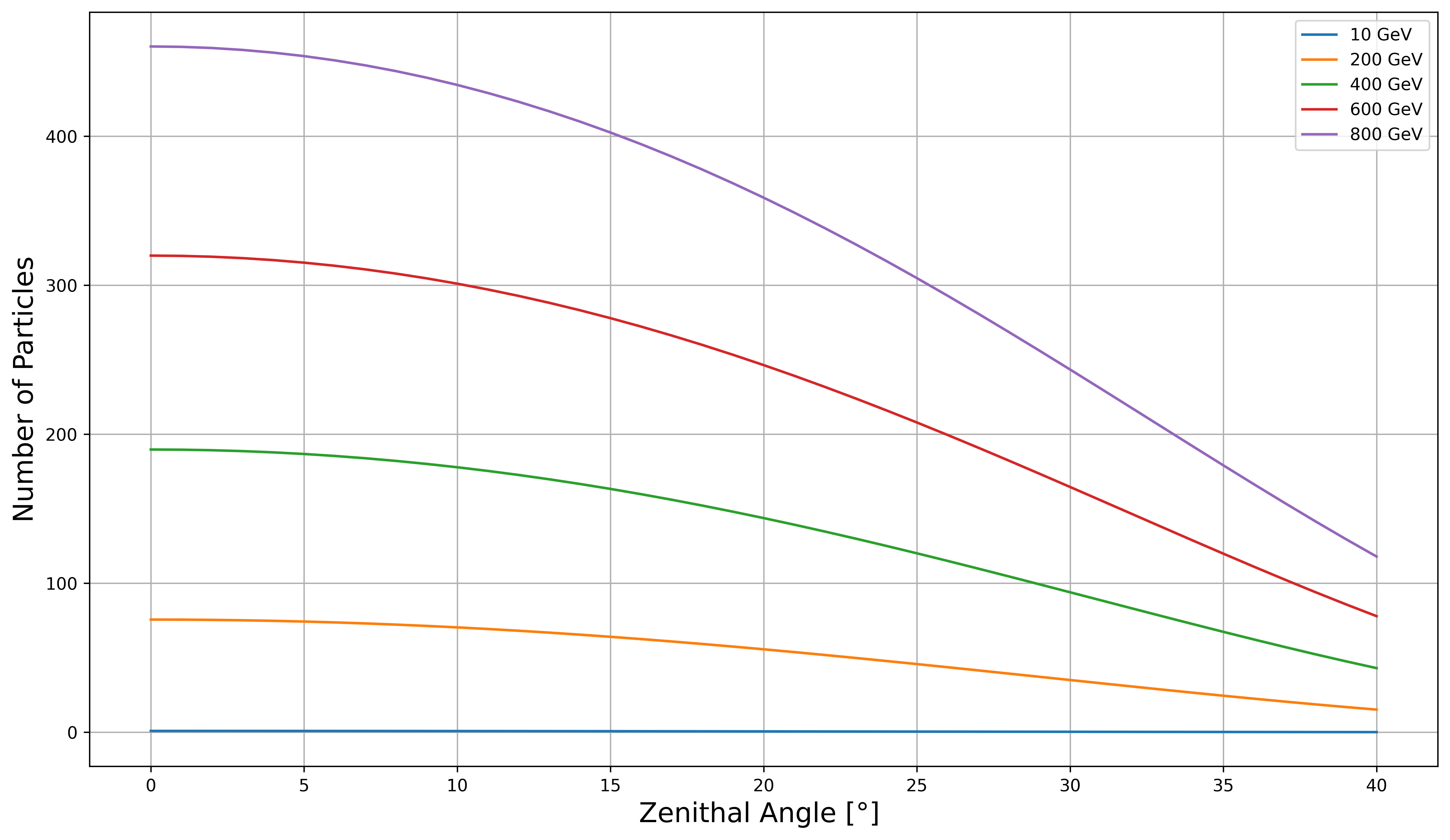}
\caption{Number of particles as a function of energy, for cosmic rays with different energies (10, 200, 400, 600, and 800 GeV) at 5300 m.}
\label{fig:angular_dependence}
\end{figure}

Figure \ref{fig:angular_dependence} illustrates the effect of the zenith angle on the number of particles $N$ for cosmic ray energy  from $10$ to $800$ GeV at a fixed altitude of $5300$ meters above sea level. These last values are motivated from the CONDOR cosmic rays observatory in Chile ~\cite{arratia2025condor}. As the zenith angle increases from $0°$ to $40°$, there is a significant decrease in the number of particles detected at a fixed energy. This behavior is a direct consequence of the increased atmospheric path length for inclined showers, which causes greater attenuation of the shower before it reaches the detector level. 

For high-altitude observatories, the number of particles detected varies significantly with the zenith angle due to the different stages of shower development observed. At high altitudes, the atmospheric depth is smaller, and inclined showers may be observed closer to their maximum development compared to vertical showers. Our formalism accurately captures this effect by properly accounting for the relationship between shower age and atmospheric depth as a function of the zenith angle.
\vspace{-0.6cm}
\section{Conclusions}

In this work, we have presented a new analytical approach to the longitudinal development of electromagnetic air showers that significantly improves the classical Greisen formalism. Our key contributions can be summarized as follows:

First, we have developed a novel analytical profile for the slope function $\lambda_1(s)$ that achieves remarkable precision with an agreement of $<0.75\%$ with the original $\lambda_1$ for shower age parameter $s$ between $0.3 < s < 1.4$. This improved function is given by:
\begin{multline}
\lambda^{1}_{MV}(s) = \frac{1}{2A}\big(1.215s - 1.215 - 3.2\ln(s) \\
+ e^{1 - s/0.1} - e^{1 - 1/0.1}\big).
\end{multline}

The introduction of the exponential term and the refinement of the coefficients from the original Greisen values provide a more accurate representation of shower development, particularly for values of $s < 1$.

We derived a complete expression for the number of particles $N(t)$ that is consistent with our improved $\lambda_1(s)$ profile:
\begin{equation}
N(E_0,t) =\frac{0.313}{\sqrt{\beta_0}}\exp\left( \lambda^1_{MV}(s)t+s\beta_0 \right).
\end{equation}

Also, we have demonstrated the consistency of our approach by deriving $\lambda_2(s)$ from our improved $\lambda_1(s)$ function and showing excellent agreement with direct calculations. This validation confirms the mathematical soundness of our formalism.

Finally, we have incorporated zenithal angle dependence into our model, making it particularly valuable for high-altitude observatories. Our analysis shows how the number of particles detected varies significantly with the zenith angle due to the increased atmospheric path length for inclined showers. Our formalism accurately captures this dependence through the transformation $t \rightarrow t/\cos\theta$ and the corresponding adjustment to the shower age parameter.

The improved analytical expressions we have developed are suitable for implementation in air shower simulation tools, offering better accuracy while maintaining computational efficiency. They are also valuable for fitting procedures over a wide range of energies and geometries, particularly for cosmic rays experiments on the ground level.

In conclusion, our new analytical approach provides a more accurate representation of electromagnetic shower development across a wider range of shower ages and zenith angles than the traditional Greisen formulation, while maintaining the mathematical tractability that makes analytical models valuable for both theoretical understanding and practical applications in cosmic ray physics.
\vspace{-0.8cm}
\begin{acknowledgments}
This work was funded by ANID PIA/APOYO AFB230003, Proyectos Internos de Investigación Multidisciplinarios USM 2024 \texttt{PI\_M\_24\_02} and Proyecto Milenio-ANID: 	\texttt{ICN2019\_044}
\end{acknowledgments}

\bibliography{apssamp}

\end{document}